# Effect of Solar-Terrestrial Phenomena on Solar Cell's Efficiency

\* Kashif Bin Zaheer, \*\* Waseem Ahmad Ansari and \*\*\*Syed Mohammad Murshid Raza

\*Mathematical Sciences Research Centre Federal Urdu University of Arts, Sciences and Technology, Gulshan-e-Iqbal Campus, Karachi-75300 (kbzaheer@fuuast.edu.pk)

\*\* Department of Mathematics University of Karachi, University Road, Karachi (waansari@uok.edu.pk)

\*\*\* Mathematical Sciences Research Centre Federal Urdu University of Arts, Sciences and Technology, Gulshan-e-Iqbal Campus, Karachi-75300 (smmurshid@fuuast.edu.pk)

## Abstract

It is assumed that the solar cell efficiency of PV device is closely related to the solar irradiance, considered the solar parameter Global Solar Irradiance (G) and the meteorological parameters like daily data of Earth Skin Temperature (E), Average Temperature (T), Relative Humidity (H) and Dew Frost Point (D), for the coastal city Karachi and a non-coastal city Jacobabad, K and J is used as a subscripts for parameters of Karachi and Jacobabad respectively. All variables used here are dependent on the location (latitude and longitude) of our stations except G. To employ ARIMA modeling, the first eighteen years data is used for modeling and forecast is done for the last five years data. In most cases results show good correlation among monthly actual and monthly forecasted values of all the predictors. Next, multiple linear regression is employed to the data obtained by ARIMA modeling and models for mean monthly observed G values are constructed. For each station, two equations are constructed the $R^2$ values are above 93% for each model, showing adequacy of the fit. Our computations show that Solar cell efficiency can be increased if better modeling for meteorological predictors governs the process.

**Nomenclature:** Global solar irradiance (G), Earth Skin Temperature (E), Average Temperature (T), Relative Humidity (H), Dew Frost Point (D), Subscripts of All Parameters taken for Karachi ($_K$), Jacobabad ($_J$), Mean Actual ($_{MA}$), Mean Forecasted ($_{MF}$), Mean Actual Karachi, Jacobabad ($_{MAK,\ MAJ}$) and Mean Forecasted Karachi Jacobabad ($_{MFK,\ MFJ}$), Multiple Linear Regression (MLR).

# 1. Introduction

The cost of traditional and other nonrenewable energy resources are on steady increase, it is essential to test, develop and depend on other alternative, non-traditional, and renewable energy sources. The most important alternative source is the solar energy. In this study, an attempt has been made to see the dependence of PV devices, which runs over solar energy, over terrestrial parameters. That is, to analyze, the efficiency of PV devices and equipment in the coastal climates particularly that of Karachi ($_K$). The results are compared with the results for non-coastal city of Sindh like Jacobabad ($_J$) which constitutes the extreme of Sindh.

The efficiency of PVC is based on different parameters which includes various geological, geometrical and electrical factors. This study establishes a statistical relationship between solar cell efficiency and some meteorological predictors having noticeable influence over the process [1]. In next section, clear evidence will be provided for the dependency of solar cell efficiency over solar irradiance. As such, an empirical relation is found between solar irradiance, taking it as response variable and some global as well as local predictors will be sought which are having an impact.

To quantify the solar radiation at any particular part of the earth's surface, position of the point, time of year, atmospheric diffusion and cloud cover, shape of the surface and reflectivity of the surface is taken in to account. However, in hilly and mountainous terrains, the distribution of slopes has major effects on surface climate and radiation amounts [9]. Surface radiation may change widely according to the Average Temperature, Relative Humidity, Dew Frost Point, Earth Skin Temperature, frequency and optical thickness of clouds, and modeling these factors successfully is important for treatment of the surface energy balance [2]. In our study, a coastal city Karachi of Sindh is taken to model the solar cell efficiency. The effect of location on the efficiency of solar cell is checked by selecting a non-coastal city Jacobabad which lie in the same region. The major difference between two cities is the location of the coast which will be analyzed for its effect of the efficiency of PVC. A comparison of the efficiency of different solar cells which are commonly used is as follows.

| Material | Level of efficiency in % Lab | Level of efficiency in % Production |
|---|---|---|
| **Mono-crystalline Silicon** | approx. 24 | 14 to17 |
| **Poly-crystalline Silicon** | approx. 18 | 13 to15 |
| **Amorphous Silicon** | approx. 13 | 5 to 7 |

## 2. Material and Method

The conversion efficiency of a photovoltaic cell array is defined as the ratio of the electrical energy produced by the array to the solar energy input to the array [7]. Expressed symbolically,

$$\eta = \frac{E}{A_c H_t \tau} \qquad (1)$$

where  $\eta$ = array efficiency

$E$ = electrical energy production of the array

$A_c$ = array area

$H_t$ = solar irradiance per unit area of the tilted array

$\tau$ = transmissivity of array cover

By the above formula of conversion efficiency of a photovoltaic cell array, the solar phenomena $H_t$ is inversely proportional to the performance of PV device. Since solar cell efficiency depends upon solar irradiance, it should be influenced by terrestrial parameters. It is assumed that the solar cell efficiency will be better in coastal climate as compare to non-coastal climate. An attempt is to be made finding a relationship between $H_t$ and different meteorological parameters. Several best suitable models are constructed using Autoregressive Integrated Moving Average (ARIMA) technique [19], to check the adequacy of actual and forecasted data, Pearson correlation as used by [4], is applied. Finally, a multiple linear regression model for Karachi ($_K$) and Jacobabad ($_J$) using all considered solar terrestrial parameters is constructed.

## 3. Results

ARIMA models depend on different p and q values, as in [5] and [11]. Modeling is applied to all the time series and a best fit model based on Minimum AICC criterion is reported. Table 1−9 shows different monthly models for each of the five parameters of Karachi ($_K$) and Jacobabad ($_J$).

For correlation, mean actual values of Global Solar Irradiance (G) is compared with mean forecasts (shown in Table 10). Time series plot is in Fig. 1 whereas the correlation b/w Mean Actual ($_{MA}$) vs. Mean Forecasted ($_{MF}$) Monthly Data of Global Solar Irradiance (G) is in Table 11. Mean Actual ($_{MA}$) and Mean Forecasted ($_{MF}$) Monthly Data for all Parameters of Karachi ($_K$) is presented in Table 12 and Fig. 2–5 represent time series plot of all the parameters for Karachi ($_K$). Table 13 shows strong correlation between Mean Actual ($_{MA}$) vs. Mean Forecasted ($_{MF}$) Monthly Data for all Parameters of Karachi ($_K$) which shows the adequacy of forecast from ARIMA. Mean Actual ($_{MA}$) and Mean Forecasted ($_{MF}$) Monthly Data for all Parameters of Jacobabad ($_J$) is presented in Table 14 and Fig. 6–9 represent time series plot of all the parameters for Jacobabad ($_J$). Table 15 shows strong correlation between Mean Actual ($_{MA}$) vs. Mean Forecasted ($_{MF}$) Monthly Data for all Parameters of Jacobabad ($_J$) which shows the adequacy of forecast from ARIMA, except correlation of $H_J$ which is weak. In Table 16, the correlation between Mean Actual ($_{MA}$) Monthly Data of G vs. all terrestrial parameters of Karachi ($_K$) and Jacobabad ($_J$) shows negative strong correlation which shows the strong inverse proportionality, whereas $H_J$ is weak. In Table 17, the correlation between Mean Forecasted ($_{MF}$) Monthly Data of G vs. all terrestrial parameters of Karachi ($_K$) and Jacobabad ($_J$) strong negative correlation which shows the strong inverse proportionality.

For multiple linear regression model, two different multiple linear regression models are constructed for Karachi ($_K$) viz., (i) modeling mean actual G versus mean of actual parameters, and, (ii) modeling mean actual G versus mean of forecasted parameters. Second equation is obtained from forecasted data from ARIMA model (last five years). $R^2$ values obtained for these models are 93.9 % and 96.4 %, showing good fitting. In the first model $D(_K)$ and $E(_K)$ shows an inverse relation whereas $H(_K)$ and $T(_K)$ shows direct relation to G. In the second model $D(_K)$, $H(_K)$ and $T(_K)$ shows an inverse relation whereas $E(_K)$ shows direct relation to G. Our result shows accurate modeling of G using the said predictors.

The MLR Model (1) is:

$$G_{MA} = 1314 - 5.44\,D_{MAK} + 1.15\,H_{MAK} - 0.60\,E_{MAK} + 3.32\,T_{MAK}$$

$$S = 5.68183 \quad R^2 = 93.9\%$$

The MLR Model (2) is

$$G_{MA} = 1400 - 0.74\,D_{MFK} - 0.553\,H_{MFK} + 1.16\,E_{MFK} - 1.17\,T_{MFK}$$

$$S = 4.36493 \quad R^2 = 96.4\%$$

Two different multiple linear regression models are constructed for Jacobabad viz., (i) modeling mean actual G versus mean of actual parameters, and, (ii) modeling mean actual G versus mean of forecasted parameters. Second equation is obtained from forecasted data from ARIMA model (last five years). $R^2$ values obtained for these models are 98.5% and 97.9%, showing good fitting. In the first model $H_J$ and $E_J$ shows an inverse relation whereas $D_J$ and $T_J$ shows direct relation to G. In the second model $D_J$, $H_J$ and $T_J$ shows an inverse relation whereas $E_J$ shows direct relation to G. Our result shows accurate modeling of G using the said predictors.

The MLR Model (1) is

$$G_{MA} = 1424 + 0.00\, D_{MAJ} - 0.102\, H_{MAJ} - 3.89\, E_{MAJ} + 2.15\, T_{MAJ}$$

$$S = 2.83633 \quad R^2 = 98.5\%$$

The MLR Model (2) is:

$$G_{MA} = 1424 - 0.020\, D_{MFJ} - 0.338\, H_{MFJ} + 0.74\, E_{MFJ} - 2.81\, T_{MFJ}$$

$$S = 3.32845 \quad R^2 = 97.9\%$$

## 4. Conclusion

This section presents the crux of all the work done in the previous section. This communication deals with a comparatively difficult task of modeling the efficiency with relation to climate. In particular, it deals with the efficiency of PVC devices in coastal as well as non-coastal stations like Karachi ($_K$) and Jacobabad ($_J$). We aim to present the efficiency of PVC in terms of G [13].

ARIMA model for Karachi ($_K$) and Jacobabad ($_J$) are constructed using first order differences to introduce stationarity in the data. For G, different monthly models are constructed and their forecast is obtained. A five year forecast (July 2000 – July 2005) considerably agrees to the data. Same is done for the remaining meteorological predictors and forecast is obtained which quite resembles the data. The correlation of mean actual versus mean forecasted monthly data of GSI is 0.773 with a $p$–value 0.003 showing sufficient statistical evidence of the ARIMA model. Correlation values among mean of each month for observed and predicted values of RH and AT, at each station shows high values except for $H_J$ whose correlation is 0.405 with $p$–value 0.191 [13].

As indicated earlier the performance and efficiency of a solar cell may depend considerably on local as well as global meteorological and geographical conditions. Thus setting G, as goal variable is sufficient enough to behave as a major factor for improvement and betterment of the efficiency and performance of PVC. Above results support our basic assumption that solar cell efficiency is directly proportional to all the terrestrial parameters taken in to account in this study. On the other hand coastal climate is more suitable for efficient performance of PVC in comparison with the non-coastal climate [13].

## 5. Tables and Graphs

Table 1: Estimated Model using first difference of Global Solar Irradiance (G)

| Month | Estimated Model | AICC |
|---|---|---|
| January | ARMA(17,3) Model | -157.107252 |
| February | ARMA(11,5) Model | -120.586942 |
| March | ARMA(7,27) Model | -233.631452 |
| April | ARMA(24,4) Model | -159.763491 |
| May | ARMA(17,5) Model | -429.347631 |
| June | ARMA(9,15) Model | -207.871830 |
| July | ARMA(7,4) Model | -233.982957 |
| August | ARMA(14,20) Model | -406.840317 |
| September | ARMA(15,11) Model | -346.512009 |
| October | ARMA(20,3) Model | -371.932621 |
| November | ARMA(24,20) Model | -361.959595 |
| December | ARMA(16,22) Model | -450.489000 |

Table 2: Estimated Model using first difference of Karachi Average Temperature ($T_K$)

| Month | Estimated Model | AICC |
|---|---|---|
| January | ARMA(20,1) Model | 1932.767830 |
| February | ARMA(7,2) Model | 1774.604427 |
| March | ARMA(26,17) Model | 1920.071916 |
| April | ARMA(7,2) Model | 1839.973945 |
| May | ARMA(11,2) Model | 1734.271557 |
| June | ARMA(10,21) Model | 1540.575483 |
| July | ARMA(5,14) Model | 1240.939277 |
| August | ARMA(8,25) Model | 1267.072572 |
| September | ARMA(6,20) Model | 1393.424217 |
| October | ARMA(4,20) Model | 1548.972269 |
| November | ARMA(26,20) Model | 1811.469225 |
| December | ARMA(17,20) Model | 1921.050659 |

Table 3: Estimated Model using first difference of Karachi Relative Humidity ($H_K$)

| Month | Estimated Model | AICC |
|---|---|---|
| January | ARMA(14,2) Model | 4167.754023 |
| February | ARMA(19,2) Model | 4025.524090 |
| March | ARMA(16,2) Model | 4277.729646 |
| April | ARMA(19,2) Model | 4005.434527 |
| May | ARMA(11,3) Model | 3960.151299 |
| June | ARMA(11,2) Model | 3533.420475 |
| July | ARMA(5,3) Model | 3426.556996 |
| August | ARMA(5,15) Model | 3311.250407 |
| September | ARMA(10,20) Model | 3562.427304 |
| October | ARMA(9,22) Model | 4059.758182 |
| November | ARMA(25,20) Model | 4017.617572 |
| December | ARMA(18,20) Model | 4051.425952 |

Table 4: Estimated Model using first difference of Karachi Earth Skin Temperature ($E_K$)

| Month | Estimated Model | AICC |
|---|---|---|
| January | ARMA(25,2) Model | 2021.120232 |
| February | ARMA(15,4) Model | 2010.085696 |
| March | ARMA(26,1) Model | 2149.125407 |
| April | ARMA(12,2) Model | 2116.314800 |
| May | ARMA(11,3) Model | 2124.932809 |
| June | ARMA(10,5) Model | 2103.872810 |
| July | ARMA(11,4) Model | 2211.646762 |
| August | ARMA(10,20) Model | 2184.130164 |
| September | ARMA(11,20) Model | 1944.307180 |
| October | ARMA(26,23) Model | 1889.963026 |
| November | ARMA(26,20) Model | 2010.910964 |
| December | ARMA(17,15) Model | 2010.710013 |

Table 5: Estimated Model using first difference of Karachi Dew Frost point ($D_K$)

| Month | Estimated Model | AICC |
|---|---|---|
| January | ARMA(11,2) Model | 3271.409557 |
| February | ARMA(14,1) Model | 3123.278906 |
| March | ARMA(8,2) Model | 3216.430011 |
| April | ARMA(24,2) Model | 2806.173789 |
| May | ARMA(11,2) Model | 2432.332673 |
| June | ARMA(11,2) Model | 1634.632513 |
| July | ARMA(12,7) Model | 1326.494099 |
| August | ARMA(9,23) Model | 1338.639200 |
| September | ARMA(25,20) Model | 1857.038296 |
| October | ARMA(9,20) Model | 2917.764646 |
| November | ARMA(19,10) Model | 3095.942204 |
| December | ARMA(18,20) Model | 3186.857201 |

Table 6: Estimated Model using first difference of Jacobabad Average Temperature ($T_J$)

| Month | Estimated Model | AICC |
|---|---|---|
| January | ARMA(24,25) Model | 2004.768936 |
| February | ARMA(5,25) Model | 2012.367522 |
| March | ARMA(11,26) Model | 2325.921771 |
| April | ARMA(26,25) Model | 2316.969978 |
| May | ARMA(10,26) Model | 2289.673543 |
| June | ARMA(12,25) Model | 2261.096701 |
| July | ARMA(15,26) Model | 2427.010984 |
| August | ARMA(5,26) Model | 2166.059007 |
| September | ARMA(7,20) Model | 1952.813770 |
| October | ARMA(26,26) Model | 2163.800188 |
| November | ARMA(26,26) Model | 2020.386352 |
| December | ARMA(11,26) Model | 2079.821741 |

Table 7: Estimated Model using first difference of Jacobabad Relative Humidity ($H_J$)

| Month | Estimated Model | AICC |
|---|---|---|
| January | ARMA(16,25) Model | 3923.931385 |
| February | ARMA(15,25) Model | 3645.387141 |
| March | ARMA(18,19) Model | 3907.447798 |
| April | ARMA(18,3) Model | 3858.129116 |
| May | ARMA(10,25) Model | 4152.879617 |
| June | ARMA(12,25) Model | 4088.057031 |
| July | ARMA(6,26) Model | 4407.651120 |
| August | ARMA(3,26) Model | 4207.524601 |
| September | ARMA(11,26) Model | 4115.625019 |
| October | ARMA(5,26) Model | 3790.872821 |
| November | ARMA(8,20) Model | 3383.676341 |
| December | ARMA(17,26) Model | 3774.797257 |

Table 8: Estimated Model using first difference of Jacobabad Earth Skin Temperature ($E_J$)

| Month | Estimated Model | AICC |
|---|---|---|
| January | ARMA(24,15) Model | 2007.753766 |
| February | ARMA(7,20) Model | 2086.863974 |
| March | ARMA(26,25) Model | 2514.856613 |
| April | ARMA(26,26) Model | 2495.379784 |
| May | ARMA(10,20) Model | 2519.256652 |
| June | ARMA(12,25) Model | 2469.849752 |
| July | ARMA(7,26) Model | 2805.295567 |
| August | ARMA(5,26) Model | 2550.541828 |
| September | ARMA(9,26) Model | 2231.568576 |
| October | ARMA(26,26) Model | 2335.724593 |
| November | ARMA(26,26) Model | 2088.521619 |
| December | ARMA(10,26) Model | 2060.863062 |

Table 9: Estimated Model using first difference of Jacobabad Dew Frost point ($D_J$)

| Month | Estimated Model | AICC |
|---|---|---|
| January | ARMA(16,25) Model | 2986.311929 |
| February | ARMA(14,25) Model | 2979.846713 |
| March | ARMA(18,26) Model | 3290.975833 |
| April | ARMA(8,26) Model | 3220.248989 |
| May | ARMA(24,13) Model | 3319.584407 |
| June | ARMA(10,25) Model | 3119.871981 |
| July | ARMA(7,26) Model | 2793.885885 |
| August | ARMA(4,26) Model | 2541.847181 |
| September | ARMA(11,26) Model | 2997.818156 |
| October | ARMA(15,19) Model | 3254.531680 |
| November | ARMA(8,20) Model | 2852.267952 |
| December | ARMA(17,24) Model | 2998.346691 |

Table 10: Mean Actual and Mean Forecasted Monthly Data of G.

| Months | Jan | Feb | Mar | Apr | May | Jun | Jul | Aug | Sep | Oct | Nov | Dec |
|---|---|---|---|---|---|---|---|---|---|---|---|---|
| $G_{MA}$ | 1390.4 | 1383.9 | 1372.0 | 1357.9 | 1345.7 | 1338.1 | 1345.7 | 1350.6 | 1358.1 | 1368 | 1380.9 | 1389.5 |
| $G_{MF}$ | 1386.9 | 1367.0 | 1365.5 | 1366.3 | 1366.5 | 1336.1 | 1366.4 | 1366.4 | 1365.8 | 1367 | 1376.5 | 1386.5 |

Table 11: Correlation b/w Mean Actual vs. Mean Forecasted Monthly Data of G.

| Parameters | Correlation | *p*-values |
|---|---|---|
| G | 0.773 | 0.003 |

Table 12: Mean Actual and Mean Forecasted Monthly Data for all Parameters of Karachi

| Months | Jan | Feb | Mar | Apr | May | Jun | Jul | Aug | Sep | Oct | Nov | Dec |
|---|---|---|---|---|---|---|---|---|---|---|---|---|
| $D_{MAK}$ | 2.745 | 5.59 | 10.996 | 15.173 | 19.441 | 22.484 | 23.315 | 22.525 | 20.634 | 14.949 | 8.405 | 2.477 |
| $H_{MAK}$ | 36.008 | 39.262 | 43.307 | 47.957 | 55.838 | 65.225 | 72.029 | 71.741 | 64.903 | 46.946 | 37.376 | 32.162 |
| $E_{MAK}$ | 20.881 | 23.984 | 29.039 | 32.087 | 33.497 | 33.67 | 32.219 | 31.426 | 31.558 | 31.557 | 27.368 | 22.22 |
| $T_{MAK}$ | 18.916 | 21.12 | 25.108 | 27.89 | 29.61 | 29.746 | 28.85 | 28.065 | 27.9 | 28.071 | 24.899 | 20.66 |
| $D_{MFK}$ | -4.947 | 5.1359 | 13.179 | 19.357 | 24.012 | 23.121 | 21.847 | 21.916 | 18.883 | 14.94 | 2.1909 | -0.828 |
| $H_{MFK}$ | 23.621 | 33.633 | 48.234 | 64.116 | 74.621 | 71.988 | 68.436 | 72.826 | 55.833 | 47.139 | 32.604 | 25.792 |
| $E_{MFK}$ | 19.102 | 29.904 | 29.017 | 30.859 | 31.349 | 30.363 | 31.386 | 30.227 | 32.467 | 31.762 | 23.773 | 21.311 |
| $T_{MFK}$ | 17.916 | 22.949 | 25.247 | 28.44 | 29.179 | 28.755 | 28.161 | 27.255 | 28.89 | 27.711 | 22.633 | 20.355 |

Table 13: Correlation between Mean Actual vs. Mean Forecasted Monthly Data for all Parameters of Karachi

| Parameters | Correlation | *p*-values |
|---|---|---|
| $D_K$ | 0.951 | 0.000 |
| $H_K$ | 0.871 | 0.000 |
| $E_K$ | 0.848 | 0.000 |
| $T_K$ | 0.961 | 0.000 |

Table 14: Mean Actual and Mean Forecasted Monthly Data for all Parameters of Jacobabad

| Months | Jan | Feb | Mar | Apr | May | Jun | Jul | Aug | Sep | Oct | Nov | Dec |
|---|---|---|---|---|---|---|---|---|---|---|---|---|
| $D_{MAJ}$ | -2.917 | -3.405 | -1.508 | 1.511 | 4.306 | 12.532 | 19.156 | 19.72 | 12.885 | 0.747 | -2.647 | -3.904 |
| $H_{MAJ}$ | 34.015 | 28.243 | 22.486 | 21.754 | 21.596 | 32.267 | 51.496 | 57.731 | 39.389 | 21.806 | 22.325 | 28.252 |
| $E_{MAJ}$ | 14.999 | 18.739 | 25.73 | 32.313 | 37.299 | 38.74 | 36.872 | 34.144 | 33.594 | 28.556 | 22.075 | 16.551 |
| $T_{MAJ}$ | 13.582 | 16.476 | 22.263 | 27.796 | 32.297 | 33.831 | 32.099 | 30.364 | 29.857 | 26.019 | 20.638 | 15.587 |
| $D_{MFJ}$ | -11.013 | -10.779 | -6.2875 | 1.728 | 10.943 | 22.519 | 14.141 | 15.926 | 3.435 | -7.548 | -10.845 | -5.8686 |
| $H_{MFJ}$ | 19.271 | 14.791 | 14.998 | 29.332 | 25.929 | 55.757 | 28.821 | 36.655 | 19.67 | 6.086 | 15.253 | 24.205 |
| $E_{MFJ}$ | 14.206 | 19.834 | 26.47 | 34.158 | 39.331 | 36.439 | 41.815 | 38.228 | 34.186 | 27.816 | 17.442 | 15.399 |
| $T_{MFJ}$ | 12.842 | 17.888 | 22.736 | 29.804 | 33.968 | 31.932 | 35.462 | 32.886 | 30.323 | 26.277 | 17.217 | 14.992 |

Table 15: Correlation b/w Mean Actual vs. Mean Forecasted Monthly Data for all Parameters of Jacobabad

| Parameters | Correlation | *p*-values |
|---|---|---|
| $D_J$ | 0.861 | 0.000 |
| $H_J$ | 0.405 | 0.191 |
| $E_J$ | 0.971 | 0.000 |
| $T_J$ | 0.975 | 0.000 |

Table 16: Correlation b/w Mean Actual Monthly Data of GSI vs. all terrestrial parameters of Karachi and Jacobabad

| Parameters | $D_{MAK}$ | $H_{MAK}$ | $E_{MAK}$ | $T_{MAK}$ | $D_{MAJ}$ | $H_{MAJ}$ | $E_{MAJ}$ | $T_{MAJ}$ |
|---|---|---|---|---|---|---|---|---|
| $G_{MA}$ | -0.964 | -0.897 | -0.925 | -0.928 | -0.812 | -0.365 | -0.989 | -0.983 |
| *p*-values | 0.000 | 0.000 | 0.000 | 0.000 | 0.001 | 0.244 | 0.000 | 0.000 |

Table 17: Correlation b/w Mean Forecasted Monthly Data of G vs. all terrestrial parameters of Karachi and Jacobabad

| Parameters | $D_{MFK}$ | $H_{MFK}$ | $E_{MFK}$ | $T_{MFK}$ | $D_{MFJ}$ | $H_{MFJ}$ | $E_{MFJ}$ | $T_{MFJ}$ |
|---|---|---|---|---|---|---|---|---|
| $G_{MF}$ | -0.743 | -0.709 | -0.699 | -0.731 | -0.696 | -0.634 | -0.654 | -0.663 |
| *p*-values | 0.006 | 0.010 | 0.011 | 0.007 | 0.012 | 0.027 | 0.021 | 0.019 |

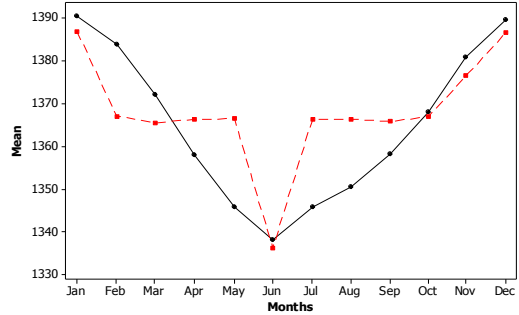

Fig. 1 Time Series Plot of $G_{MA}$ vs. $G_{MF}$

Fig. 2–5: Time Series Plot For Mean Actual vs. Mean Forecasted Monthly Data of Karachi.

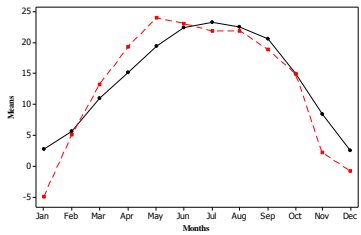 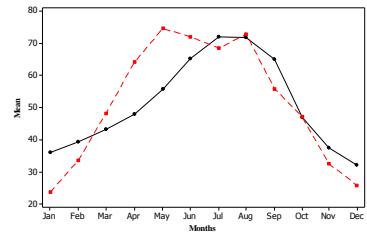

Fig. 2 Time Series Plot of Mean $D_K$  Fig. 3 Time Series Plot of Mean $H_K$

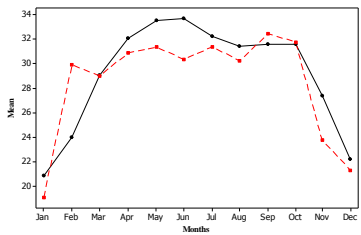 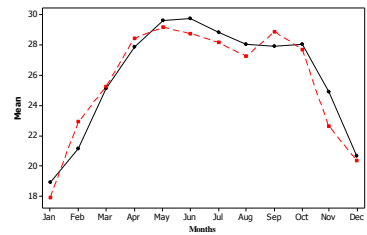

Fig. 4 Time Series Plot of Mean $E_K$  Fig. 5 Time Series Plot of Mean $T_K$

Fig. 6–9: Time Series Plot For Mean Actual vs. Mean Forecasted Monthly Data of Jacobabad.

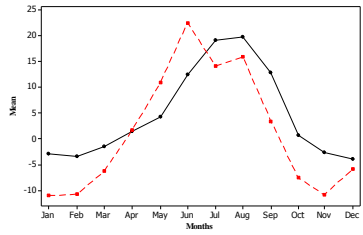

Fig. 6 Time Series Plot of Mean $D_J$

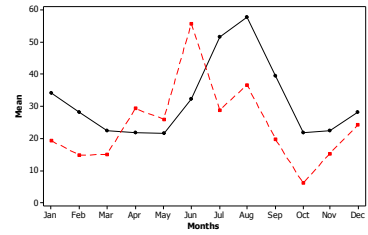

Fig. 7 Time Series Plot of Mean $H_J$

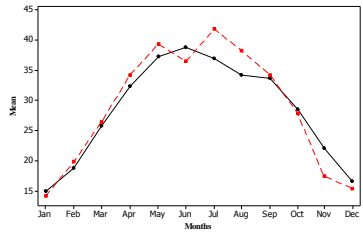

Fig. 8 Time Series Plot of Mean $E_J$

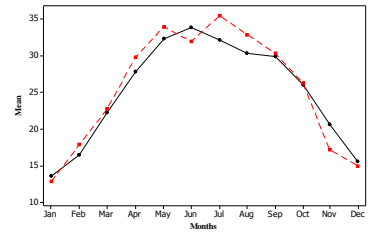

Fig. 9 Time Series Plot of Mean $T_J$